\documentclass{PoS}

\usepackage{epsf}
\usepackage{amssymb}
\usepackage{amsmath}
\usepackage{amsfonts}
\usepackage{cite}
\usepackage[small]{caption2}

\usepackage[T1]{fontenc}
\usepackage{psfrag,epsfig,graphicx,graphics}

\def\eps{\epsilon}

\newcommand{\f}{\frac}
\newcommand{\be}{\begin{equation}}
\newcommand{\beq}{\begin{equation}}
\newcommand{\ee}{\end{equation}}
\newcommand{\eq}{\end{equation}}
\newcommand{\eeq}{\end{equation}}

\newcommand{\bea}{\begin{eqnarray}}
\newcommand{\eea}{\end{eqnarray}}

\def\slashchar#1{\setbox0=\hbox{$#1$}
   \dimen0=\wd0
   \setbox1=\hbox{/} \dimen1=\wd1
   \ifdim\dimen0>\dimen1
      \rlap{\hbox to \dimen0{\hfil/\hfil}}
      #1
   \else
      \rlap{\hbox to \dimen1{\hfil$#1$\hfil}}
      /gdatdafinal2.tex
   \fi}

\newcommand{\alp}{\alpha '}
\def\ba{\begin{eqnarray}}
\def\ea{\end{eqnarray}} 
 \def\bi{\begin{itemize}}
 \def\ei{\end{itemize}}
 
 \def\ka{\kappa}
\newcommand{\la}{\label}

\newcommand{\va}{\varphi} 
\newcommand{\bv}{\bar \varphi} 

\title{Traveling wave solution of the Reggeon Field Theory}

\ShortTitle{Traveling wave solution of the Reggeon Field Theory}

\author{
\speaker
{Robi~Peschanski}\\
Institut de Physique Th{\'e}orique and URA 2306, Unit\'{e} de Recherche associ{\'e}e au CNRS, CEA-Saclay, F-91191 Gif/Yvette Cedex, France\\
E-mail: \email{robi.peschanski@cea.fr}}

\abstract{We identify the nonlinear evolution equation in impact-parameter space
 for the ``Supercritical Pomeron'' in Reggeon Field Theory as a 2-dimensional
  stochastic Fisher and Kolmogorov-Petrovski-Piscounov equation. It exactly preserves 
  unitarity and leads in its radial form to an high energy traveling wave 
  solution corresponding to an ``universal'' behaviour of the impact-parameter 
  front profile of the elastic amplitude; Its rapidity dependence and form 
  depend only on one parameter, the noise strength, independently of the initial
   conditions and of the non-linear terms restoring unitarity. Theoretical 
   predictions are presented for the three typical different regimes
    corresponding to zero, weak and strong noise, respectively. They have phenomenological 
    implications for total and differential hadronic cross-sections at colliders.
}

\FullConference{European Physical Society Europhysics Conference on High Energy Physics,
EPS-HEP 2009,\\
		 July 16 - 22 2009\\
		 Krakow, Poland}

\begin{document}

\section{From Reggeon Field Theory (1967-1975) to  the stochastic sFKPP
 Equation}
\label{Sec_Int}

Reggeon Field Theory (RFT) has been motivated, 
long ago, by an effective field-theoretical treatment \cite{Mald} of the Strong 
Interaction physics at high-energy, 
commonly called nowadays ``soft interactions'', since they involved only rather small
 transfer of momentum. As such, they belong to the strongly-coupled regime of QCD
 and thus not reducible to perturbative calculations. The field-theoretical action
 is \cite{Mald}
 \begin{eqnarray}
\label{Action}
S[\bv,\va]=\f 1\alp \int d^2b\ dY\ 
\left\{{\bv\left[\partial_Y-\alp\nabla^2\right]\va-\mu\bv\va}
+{{g\left(\bv\va^2-\ka\bv^2\va\right)}}
\right\} \, ,
\end{eqnarray}
where  $\va(Y,\vec b)$ is the Reggeon field. It is considered as a non-relativistic 
quantum field moving 
in a (1+2)-dimensional space spanned by $Y,$ the total rapidity considered 
as a time variable and  $\vec b$ the impact-parameter vector in the transverse 
plane
 of the interaction. $\alp$ is the Regge slope, $\mu+1$ is the
  Reggeon intercept. The coupling constants  $g$ and $g\ka$
  characterize the Reggeon self interactions, such that $g$ is the coupling strength for the 
  $2\to 1\ merging$ while
 $g\kappa$ defines the $1\to 2\ splitting$ vertex. Note that  $\kappa=1$
  recovers the original symmetric  RFT model \cite{Mald}. The 
 field-theoretical formalism
  relies on the introduction of  an auxiliary reggeon field $\bv$ which plays
   the role of a source term.
  It is important to realize that the theory is nonhermitian 
  ($\bv \ne \va\dag $)
  which is at the root of a deep relation with out-of-equilibrium statistical physics of
   reaction-diffusion type, 
  as we shall see.
 
We are interested in the so-called ``Supercritical'' Reggeon problem where the
Born term (corresponding to retaining only the quadratic part of \eqref{Action})
reads
\begin{eqnarray}
\label{Pom}
\va(Y,\vec b)=\exp{\mu Y-\f {b^2}{4\alp Y}}\ ;\quad  {\mu > 0}  ,
\end{eqnarray}
which has been shown to be phenomenologically preferred but for which no 
theoretical solution has yet been found for the corresponding RFT. 
Our goal is to
show that, using an intimate connection with  statistical physics one is able 
\cite{Jani} to 
find the solution of the RFT giverned by the action \eqref{Action}.

Using  the Stratonovitch transformation 
to linearize the action as a function of the auxiliary field $\bv$
and performing the  functional integral 
$\int {\cal D}\bv {\cal D}\va\ \exp{\left(\f i\alp S[\bv,\va]\right)
}$ one gets \cite{Jani} a nonlinear 
Langevin equation for the Reggeon solution $\va(Y,\vec b)$. After reparametrisation,
 the equation boils down to
\ba
\partial_t \va(t,\vec r)= \nabla^2_r \va+\va-\va^2+ \eps\sqrt{\va\left(1-\va\right)}\
 \nu\left(t,\vec r\right)\ ,
\la{Langevin}
\ea
where $\nu\left(t,\vec r\right)$ is the two-dimensional white noise. 
The dictionnary between the RFT and sFKPP is
\ba
{Time:}\ t=\mu (Y\!-Y_0)\ ;\quad {Space:}\ \vec r =  \f \mu \alp \vec b\ ; \quad 
{Noise:}\ 
 \eps = \sqrt{2\mu \ka}\ .
\la{dico}
\ea
Eq.\eqref{Langevin}
 is nothing else than the extension in    2 dimensions of the known 
 stochastic Fisher and 
 Kolmogorov, Petrovsky, Piscounov equation (sFKPP).

In this way, the RFT is mapped onto  an equation describing non-equilibrium
 processes   in a  2-dimensional
 space with 
 diffusion, creation, merging and splitting as a function of time. In 
 Eq.\eqref{Langevin}, all coupling strengths characterizing these processes 
 can be  absorbed in the redefinitions \eqref{dico}, except for
 the parameter $\eps=\sqrt{2\mu \ka}$ which thus appears as the only free parameter. 
  
\section{Traveling waves and the RFT ``Phase Diagram''}
\la{Sec-Early}

In the problem of deriving solutions for the  azimuthally symmetric
 Reggeon amplitude, the 2-dimensional 
problem can be further reduced to a radial 
equation, with $r\equiv \vert\vec r\vert$ 
\ba
\partial_t \va(t,r)= \partial_{rr} \va+ {\f 1r \partial_r \va }
+\va-\va^2+\eps\sqrt{\f{\va\left(1-\va\right)}{{2\pi r}}}\
 \nu\left(t,r\right)\ .
 \la{radial}
 \ea
We are thus led to consider  asymptotic solutions of the 1-d {\it radial}
sFKPP equation.

 The main property of the 1d sFKPP equations \cite{FKPP} is the 
existence of asymptotic {\it traveling wave} solutions $\va(t,r) \to \va
 \left(r-r(t)\right)$
characterized by
 a wave front $r(t) \sim vt$ where the solution jumps from the value
  $1$ to $0$ and moving with the speed $v.$ Apart from the ${\f 1r \partial_r \va }$ term which is 
subdominant at large $r,$   the main change w.r.t. the standard 1d sFKPP
equation is in the noise term which is now characterized by an effective coupling strength
\ba
\zeta \sim \f {\eps}{\sqrt{2\pi r(t)}}
= \ \sqrt{\f{\ka\sqrt{\alp\mu}}{2\pi b_s(Y)}}\ ,
\la{noise}
\ea
where we have approximated the $r$-dependence of the noise by its value at the 
wave front (where $\va\left(1-\va\right)$ is not small) and 
$b_s(Y)\equiv\alp/\mu r(t)$ is the effective impact-parameter radius. Note
 that the 
noise \eqref{noise} is reduced by a factor of order $\sqrt(vt)$ compared to 
 the 2d-sFKPP equation due to the
azimuthal symmetry, since it has to be restricted to  azimuthally
 symmetric fluctuations. 
 
\begin{figure}[b]
  \includegraphics[width=8cm]{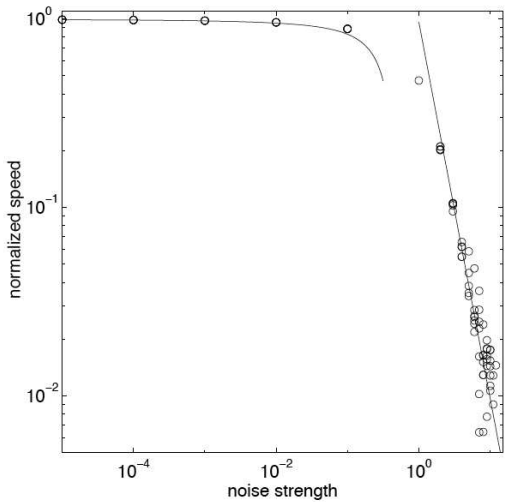}
  \caption{Normalized  wave speed as a function of
   the normalized noise strength (from Ref.\cite{Beuf}).\newline
   White dots: numerics; Lines:
   analytic results.}
\label{fig1}
\end{figure}

 The typical structure of the solutions for the wave speed is  displayed  (for the 1d sFKPP
  equation \cite{Beuf}) in Fig.\ref{fig1}. The vertical axis is the 
  speed of the wave front $v/v_0$ normalized by the no-noise speed
   $v_0\equiv 2.$ The horizontal axis is the normalized noise strength, represented by
   $\zeta$ in the approximation \eqref{noise}. One is led to distinguish mainly
   three different regimes of traveling wave solutions which may play 
   the role of a RFT phase diagram depending on the noise strength. 

Let us restore the appropriate 
high-energy observables, namely $b_s(Y),$ the  impact-parameter radius, 
$\sigma_{tot} \propto b^2_s(Y),$
the total cross-section, and $T(Y,b),$ the impact-parameter dependence of the 
imaginary elastic amplitude and discuss the three different regimes appearing
 in Fig.\ref{fig1}.

\noindent The first ``phase'' is the no-noise regime where  $\zeta \sim \ka \approx 0,$  
$i.e.$
when {\it splitting $\ll$ merging}. One gets, 
\ba
\sigma_{tot}\propto b^2_s(Y)=\left\{b_0 + 
2\sqrt{\alp\mu}
\left[Y-Y_0-\f 1\mu\ \log {\f Y{Y_0}}+\cdots \right]\right\}^2 ; \quad
T(Y,b) \sim T(b-b_s(Y))\ . 
\la{no-noise}\ea
The second phase is the so-called ``weak-noise'' phase where $\ka \approx 
{\cal O}(1),$ and {\it splitting $\sim$ merging}, corresponding to the original RFT
action \eqref{Action} and leading to
\ba
\sigma_{tot} =\left\{b_0 + 2\sqrt{\alp\mu}\ (Y-Y_0)\left[1-\f {\pi^2}
{2\log^2\left(2{\zeta}^{-2}\right)}\right]\right\}^2 \ ;\
T(Y,b) \sim T\left(\f {b-b_s(Y)}{D\sqrt{\alp(Y-Y_0)}}\right) .
\la{weak-noise}\ea
The third phase, corresponding to ``strong noise'' with $\ka 
\approx {\cal O}(100),$  and  {\it splitting $\gg$ merging} gives
\ba
\sigma_{tot} = b^2_s(Y_0)\ \exp{\left[\f{8\pi}\ka(Y-Y_0)\right]}\ ;   \quad
T(Y,b) \sim {\rm erfc}\left(\f {b-b_s(Y)}{D\sqrt{\alp(Y-Y_0)}}\right)\ . 
 \la{strong noise}\ea

It is interesting to note that, due to the energy dependence $\zeta \propto 
 1/{\sqrt Y},$ see \eqref{noise}, there is a natural evolution from strong to 
weak and ultimately no-noise ``phases'' as a function of energy. Hence, for
 instance,  the  apparently exponential growth of the cross-section \eqref{strong noise} at 
 strong noise, which would violate the Froissart bound, is tamed by
  the noise strength 
 evolution transfering the energy behaviour into the $Y^2$ behaviour of 
 \eqref{weak-noise} and finally ending into \eqref{no-noise}.

Let us summarize the main consequences of our study \cite{Jani}:
\begin{itemize}
\item The  supercritical Reggeon Field Theory  can be 
exactly mapped onto the  2d stochastic Fisher and Kolmogorov-Petrovski-Piscounov 
equation.
\item  {The solution for the elastic amplitude at high-energy
 is given by  universal traveling wave solutions of the radial sFKPP equation.}
\item {A ``Phase Diagram'' depending only on the } ratio between the
  {\it splitting} over {\it merging} coupling of the Reggeons plays the role 
  of a dynamical ``order parameter'', characterizing the solution.
\end{itemize}

It is to be remarked that the asymptotic behaviour does not 
depend on the rather strong renormalization of the ``supercritical'' Born term 
\eqref{Pom}. Thinking in terms of Regge singularities, such universality, observed
 experimentally in hadron-hadron reactions and  usually
 associated to the Regge pole structure of the Born term \eqref{Pom}, is here 
 restored through the universality of traveling wave solutions while the Born term
 is completely masked  by  Reggeon cuts through unitarity restoration.
The final word is  now to be given to phenomenology.


\begin{thebibliography}{99}





\bibitem{Mald}
  H.~D.~I.~Abarbanel, J.~B.~Bronzan, R.~L.~Sugar and A.~R.~White,
  Phys.\ Rept.\  {\bf 21}, 119 (1975).

 \bibitem{Jani}
  R.~Peschanski,
  Phys.\ Rev.\  D {\bf 79}, 105014 (2009).

\bibitem{FKPP}
R.~A. Fisher,
Ann. Eugenics {\bf 7}, 355 (1937);
A.~Kolmogorov, I.~Petrovsky, and N.~Piscounov,
Moscou Univ. Bull. Math. {\bf A1}, 1 (1937).

\bibitem{Beuf}
  C.~R.~Doering, C,~Mueller and P.~Smereka,
  Phys.\ {\bf A 325}, 243 (2003). 
  
\end{thebibliography}
\end{document}